\input harvmac
\input epsf
\input psfig
\hfuzz=100pt

\overfullrule=0pt
\parindent=0pt


\def\G(#1){\Gamma(#1)}

\def\C|#1{{\cal #1}}
\def\(#1#2){(\zeta_#1\cdot\zeta_#2)}
\def\lr{\lref}


\def\xxx#1 {{hep-th/#1}}
\def\lr { \lref}
\def\npb#1(#2)#3 { Nucl. Phys. {\bf B#1} (#2) #3 }
\def\rep#1(#2)#3 { Phys. Rept.{\bf #1} (#2) #3 }
\def\plb#1(#2)#3{Phys. Lett. {\bf #1B} (#2) #3}
\def\prl#1(#2)#3{Phys. Rev. Lett.{\bf #1} (#2) #3}
\def\physrev#1(#2)#3{Phys. Rev. {\bf D#1} (#2) #3}
\def\ap#1(#2)#3{Ann. Phys. {\bf #1} (#2) #3}
\def\rmp#1(#2)#3{Rev. Mod. Phys. {\bf #1} (#2) #3}
\def\cmp#1(#2)#3{Comm. Math. Phys. {\bf #1} (#2) #3}
\def\mpl#1(#2)#3{Mod. Phys. Lett. {\bf #1} (#2) #3}
\def\ijmp#1(#2)#3{Int. J. Mod. Phys. {\bf A#1} (#2) #3}

\def\lam16{\lambda^{16}}
\def\simlt{\mathrel{\lower2.5pt\vbox{\lineskip=0pt\baselineskip=0pt
           \hbox{$<$}\hbox{$\sim$}}}}
\def\simgt{\mathrel{\lower2.5pt\vbox{\lineskip=0pt\baselineskip=0pt
           \hbox{$>$}\hbox{$\sim$}}}}

\parindent 25pt
\overfullrule=0pt
\tolerance=10000

\sequentialequations

\noblackbox
\baselineskip 12pt plus 2pt minus 2pt

\Title{\vbox{\baselineskip12pt
\hbox{hep-th/9812093}
\hbox{CPTH-S693.1298}
\hbox{LPTENS-98/44 }
}}
{\vbox{\centerline{Branes and the Gauge Hierarchy}
}}
\bigskip
\centerline{Ignatios Antoniadis }
\medskip
\centerline{  Centre de Physique Th{\'e}orique (CNRS UMR 7644)}
\centerline{Ecole Polytechnique, 91128 Palaiseau, France}
\centerline{\it antoniad@cpht.polytechnique.fr}
\medskip
\centerline{and}
\medskip
{\centerline{Constantin Bachas}}
\medskip
\centerline{Laboratoire de Physique Th{\'e}orique (CNRS UPR 701)}
\centerline{ Ecole Normale Sup{\'e}rieure, 75231 Paris Cedex, France}
\centerline{\it  bachas@physique.ens.fr}
 \bigskip

\centerline{\bf Abstract}
 \bigskip

  If the fundamental type-I string scale is of the order of few  TeV,
 the problem of the gauge
  hierarchy is that of understanding why  some dimensions transverse to
  our brane-world are so large. The {\it technical}  aspect of this
  problem, as usually formulated, is  `why  quantum corrections do not
  modify drastically the masses and other parameters of the Standard
  Model'. We argue  that within type-I  perturbation theory, the
  technical hierarchy problem is  solved (a) if all massless tadpoles cancel
  locally over distances  of order  the string length
in the transverse space,
or (b) if the massless fields with uncancelled local tadpoles propagate
`effectively'
 in  $d_\perp \ge 2 $ large transverse dimensions.
These restrictions ensure that
  loop corrections to the Standard Model parameters
  decouple   from the four-dimensional
  Planck   scale, except when there are uncancelled tadpoles in $d_\perp =2$
 in which case  the dependence on $M_P$ is  logarithmic.  This
  latter case is thus singled out as the only one in  which the origin of the
  hierarchy would not be  attributed entirely to `out of this world'
  bulk physics. The role of the renormalization group equations in
  summing the leading large logs is replaced by the classical
  2d supergravity equations in  the transverse space.

\bigskip
\vskip 0.8cm

\Date{December  1998}
 \vfill\eject
\lr\brane{J. Polchinski, {\it Dirichlet-Branes and Ramond-Ramond Charges},
Phys. Rev. Lett. {\bf 75} (1995) 4724, hep-th/9510017.}
\lr\revs{
J. Polchinski, {\it TASI Lectures on D-branes}, hep-th/9611050; 
C. Bachas, {\it Lectures on D-branes}, hep-th/9806199.}
\lr\horwitt{
P. Horava and Witten, {\it Heterotic and Type I String Dynamics from Eleven
Dimensions},  Nucl. Phys. {\bf B460} (1996) 506, hep-th/9510209.}
\lr\addtwo{N. Arkani-Hamed, S. Dimopoulos and G. Dvali,
{\it Phenomenology, Astrophysics and Cosmology of Theories with Sub-millimeter
Dimensions and TeV Scale Quantum Gravity},
hep-ph/9807344.}
\lr\giu{ G.F. Giudice, R. Rattazzi and J.D. Wells,
{\it Quantum Gravity and Extra Dimensions at High-Energy Colliders},
hep-th/9811291;
E.A. Mirabelli, M. Perelstein and M.E. Peskin,
{\it Collider Signatures of New Large Space Dimensions},
hep-ph/9811337;
J. L. Hewett, {\it Indirect Collider Signals for Extra Dimensions}, 
hep-ph/9811356;
P. Mathews, S. Raychaudhuri and  K. Sridhar, {\it Getting to the top with extra dimensions},
hep-ph/9811501.
 }
\lr\berkooz{M. Berkooz, R. Leigh, J. Polchinski, J. Schwarz,
N. Seiberg, E. Witten,
{\it Anomalies, Dualities, and Topology of D=6 N=1 Superstring Vacua},
Nucl. Phys. {\bf B475} (1996) 115, hep-th/9605184.}
\lr\add{N. Arkani-Hamed, S. Dimopoulos and G. Dvali,
{\it The Hierarchy Problem and New Dimensions at a Millimeter},
Phys. Lett. {\bf B429} (1998) 263, hep-ph/9803315.}
\lr\aadd{I. Antoniadis, N. Arkani-Hamed, S. Dimopoulos and G.
Dvali, {\it New Dimensions at a Millimeter to a Fermi and Superstrings at a
TeV},
Phys. Lett. {\bf B436} (1998) 263, hep-ph/9804398.}
\lr\st{G. Shiu and S.-H.H. Tye,
{\it TeV Scale Superstrings and Extra Dimensions},
Phys. Rev. {\bf D58} (1998) 106007, hep-th/9805157;
Z. Kakushadze and S.-H.H. Tye, {\it Brane World}, hep-th/9809147.}
\lr\ia{I. Antoniadis,
{\it A Possible New Dimension at a Few TeV},
Phys. Lett. {\bf B246} (1990) 377.}
\lr\b{C. Bachas (1995), unpublished;
K. Benakli, {\it Phenomenology of Low Quantum Gravity Scale Models},
hep-ph/9809582.}
\lr\ly{J.D. Lykken,
{\it Weak Scale Superstrings}, Phys. Rev. {\bf D54} (1996) 3693,
hep-th/9603133.}
\lr\adpq{I. Antoniadis, S. Dimopoulos, A. Pomarol and M. Quir\'os,
{\it Soft Masses in Theories with Supersymmetry Breaking by
TeV-Compactification},
hep-ph/9810410.}
\lr\ads{I. Antoniadis, E. Dudas and A. Sagnotti,
{\it Supersymmetry Breaking, Open Strings and M-theory}, hep-th/9807011.}
\lr\tfour{I. Antoniadis, C. M\~unoz and M. Quir\'os,
{\it Dynamical Supersymmetry Breaking With a Large Internal Dimension},
Nucl. Phys. {\bf B397} (1993) 515, hep-ph/9211309;
I. Antoniadis, S. Dimopoulos and G. Dvali,
{\it Millimetre-range Forces in Superstring Theories with Weak-scale
Compactification}, Nucl. Phys. {\bf B516} (1998) 70, hep-ph/9710204.}
\lr\str{S. Ferrara, C. Kounnas and F. Zwirner,
{\it Mass Formulae and Natural Hierarchy in String Effective Supergravities},
Nucl. Phys. {\bf B429} (1994) 589, hep-th/9405188.}
\lr\ff{C. Bachas, C. Fabre, E. Kiritsis, N. Obers and P. Vanhove,
{\it Heterotic/Type I Duality and D-brane instantons},
Nucl. Phys. {\bf B509} (1998) 33, hep-th/9707126;
E. Kiritsis and N. Obers,
{\it Heterotic-Type I Duality in $d<$10 Dimensions, Threshold Corrections and
D-instantons}, JHEP {\bf 10} (1997) 4, hep-th/9709058;
W. Lerche and S. Stieberger,
{\it Prepotential, Mirror Map and F-theory on K3}, hep-th/9804176;
M. Bianchi, E. Gava, F. Morales and K.S. Narain,
{\it D-strings in Unconventional Type I Vacuum Configurations}, hep-th/9811013;
W. Lerche, S. Stieberger and N. Warner,
{\it Quartic Gauge Couplings from K3 Geometry}, hep-th/9811228.}
\lr\stab{Sundrum,
{\it Compactification For a Three-Brane Universe}, hep-ph/9805471;
K.R. Dienes, E. Dudas, T. Gherghetta and A. Riotto,
{\it Cosmological Phase Transitions and Radius Stabilization in Higher
Dimensions},
hep-ph/9809406;
N. Arkani-Hamed, S. Dimopoulos and J. March-Russell,
{\it Stabilization of Submillimeter Dimensions: The New Guise of the Hierarchy
Problem}, hep-th/9809124.}
\lr\gk{For a review see A. Giveon and Kutasov,
{\it Brane Dynamics and Gauge Theory}, hep-th/9802067.}
\lr\bs{M. Bianchi and A. Sagnotti,
{\it On the Systematics of Open String Theories},
Phys. Lett. {\bf B247} (1990) 517;
{\it Twist Symmetry and Open String Theories},
Nucl. Phys. {\bf B361} (1991) 519;
 E. Gimon and J. Polchinski,
{\it Consistency Conditions for Orientifolds and D Manifolds},
 Phys. Rev. {\bf D54}  (1996) 1667, hep-th/9601038.}
\lr\kleb{M.R. Garousi and R.C. Myers,
{\it Suprstring Scattering from D-branes},
Nucl. Phys. {\bf B475} (1996) 193, hep-th/9603194;
A. Hashimoto and I.R. Klebanov, {\it Decay of Excited D-branes},
Phys. Lett. {\bf B381} (1996) 437, hep-th/9604065.}
\lr\cb{C. Bachas, {\it Unification with Low String Scale},
hep-ph/9807415, to appear in JHEP.}
\lr\karim{I. Antoniadis and K. Benakli,
{\it Limits on Extra Dimensions in Orbifold Compactifications of Superstrings}
Phys. Lett. {\bf B326} (1994) 69, hep-th/9310151.}
\lr\F{C. Vafa, {\it Evidence for F Theory},
Nucl. Phys. {\bf B469} (1996) 403, hep-th/9602022;
 A. Sen, {\it F-theory and Orientifolds},
Nucl. Phys. {\bf B475} (1996) 562, hep-th/9605150.}
\lr\seven{E. Bergshoeff, M. de Roo, M. B. Green, G. Papadopoulos,
 P. K. Townsend, {\it Duality of Type II 7-branes and 8-branes},
 Nucl. Phys. {\bf B470} (1996) 113, hep-th/9601150.}
\lr\cyril{C. Bachas and C. Fabre,
{\it Threshold Effects in Open-String Theory},
 Nucl. Phys. {\bf B476} (1996) 418, hep-th/9605028;
I. Antoniadis, H. Partouche and T.R. Taylor,
{\it Duality of N=2 Heterotic-Type I Compactifications in Four Dimensions},
Nucl. Phys. {\bf B499} (1997) 29, hep-th/9703076.}
\lr\w{E. Witten,
{\it Strong Coupling Expansion of Calabi-Yau Compactification},
Nucl. Phys. {\bf B471} (1996) 135, hep-th/9602070.}
\lr\power{
K.R. Dienes, E. Dudas and T. Gherghetta,
{\it Extra Space-Time Dimensions and Unification},
Phys. Lett. {\bf B436} (1998) 55, hep-ph/9803466;
{\it Grand Unification at Intermediate Mass Scales Through Extra Dimensions},
hep-ph/9806292;
Z. Kakushadze, {\it Novel Extension of MSSM and `TeV Scale' Coupling
Unification}, hep-th/9811193.}
\lr\coupl{R. Sundrum,
{\it Effective Field Theory for a Three-Brane Universe}, hep-ph/9805471;
T.Han, J.D. Lykken and R.-J. Zhang,
{\it On Kaluza-Klein States From Large Extra Dimensions}, hep-ph/9811350.}
\lr\polwit{J. Polchinski and E. Witten, {\it 
Evidence for Heterotic - Type I String Duality}, Nucl.Phys. {\bf B460} (1996) 525,
hep-th/9510169.}
\lr\mag{C. Bachas, {\it A Way to Break Supersymmetry}, hep-th/9503030;
{\it Magnetic Supersymmetry breaking}, hep-th/9509067.}
\lr\doug{M. Berkooz, M. Douglas and  R. Leigh, {\it Branes Intersecting at Angles},
Nucl.Phys. {\bf B480} (1996) 265, 
 hep-th/9606139.}
\lr\bgs{C. Bachas, M.B. Green and A. Schwimmer, {\it
 (8,0) Quantum mechanics and symmetry enhancement in type I' superstrings}, 
J.High Energy Phys. 01 (1998) 006, 
 hep-th/9712086.}
\lr\size{ M.R.  Douglas, D. Kabat, P. Pouliot and  S.H. Shenker,
{\it D-branes and Short Distances in String Theory},  Nucl.Phys. {\bf B485} (1997) 85,
hep-th/9608024.}


\noblackbox
\baselineskip 14pt plus 2pt minus 2pt

  The effective Lagrangian describing low-energy physics in our world
is
\eqn\sm{
S_{eff} = \int d^4 x\sqrt{g}  \left( \Lambda  +  M_P^2  {\cal  R} +
{\cal L}_{SM} \right),
}
where ${\cal L}_{SM}$ is the  Lagrangian for the  gauge,
matter and Higgs fields of the Standard Model. The gauge hierarchy and
cosmological constant problems are the questions of why, compared to
the typical  Standard Model scale around the TeV,
 the  two  scales of  gravity,
 $M_P \sim 10^{15}$ TeV
and  $\Lambda^{1/4} \simlt 10^{-15}$ TeV respectively,
are so hierarchically
different. There are of course some extra  hierarchies in  the
parameters of the Standard Model per se, mainly in connection to
fermion masses. We will however adopt  the point of view that
  at least the most
severe one --  the apparent smallness of neutrino masses --  is
intimately related to the basic gauge hierarchy,
and will prove to be natural once this latter has been explained.

The {\it technical} aspect of the gauge-hierarchy problem is a
question of more limited scope. The question is why   radiative
corrections do not destabilize the masses and other parameters of the
Standard Model. In the context of  renormalizable local field theory,
spontaneously or softly-broken supersymmetry solves this  problem by
allowing only
logarithmic dependence on the ultraviolet cutoff,
usually taken at or near $M_P$.

More recently it has been proposed
that the fundamental scale $M_s= {1/l_s}$ of string theory could
 be in the TeV region {\ly}, with
the Standard Model fields living on what looks at distances larger
than $l_s$ as a three-brane {\brane \horwitt}.  The weakness of
gravity  would then be attributed  to the existence of  large
transverse dimensions, in the submillimeter to the subfermi region, in
which only the gravitational sector propagates \add.
 This possibility has a perturbative
description within  type-I string theory, which
necessarily implies some submillimiter dimensions when the fundamental
tension is near the TeV \aadd.\foot{Earlier efforts
 to lower the  compactification {\ia}  or
string scale {\b}   near  the TeV region have been  made  in the context of the
heterotic string, while the possibility of lowering the unification
scale has been discussed more recently in {\power}.}
Furthermore, explicit model
building looks promising \st, while this radical proposal appears to be
a priori compatible
with all experimental constraints if the number of  large transverse
dimensions is larger than two, and marginally compatible if it is 
equal to two \addtwo.

One would think that in such a
scenario  the technical aspect of the gauge hierarchy problem is
automatically solved, since the ultraviolet cutoff $M_s$
is  of  the same order as  the electroweak  scale \add.  It was pointed out,
however, in ref. {\cb}  that within type-I perturbation theory
gauge couplings and other parameters of the world-brane theory could
receive corrections growing linearly or logarithmically
 with the size $R$ of the transverse space, and hence a posteriori with
$M_P$. The precise rate of growth was attributed to the
`effective dimensionality'  of the transverse space. Since such corrections
would  modify the parameters of the Standard Model at the electroweak scale,
controlling them can be considered as the novel guise of
the technical problem associated with  the gauge hierarchy.

In this letter we further elucidate the origin of these corrections, and
state the conditions under which the technical hierarchy problem can
be  solved. As we will explain, such
effects will be  completely absent  (a) if all tadpoles of
massless bulk fields cancel locally
at scales $>l_s$ in transverse space, or (b) if bulk fields with
uncancelled local tadpoles propagate `effectively' in $d_\perp>2$
large transverse dimensions. 
The limiting case $d_\perp= 2$ leads to
logarithmic sensitivity of the world-brane parameters
on $R$ and $M_P$, while for $d_\perp=1$ the
sensitivity is linear  and one  hits a strong-coupling
singularity {\polwit \w}
 much before $M_P$ is allowed  to reach 
its experimentally-determined  value. It is important to stress that $d_\perp$
need not be the total number of superstringy  dimensions for two
reasons {\cb} :  because linear divergences may  persist when one transverse dimension
grows much larger than all the others, and  because some of the
tadpoles may couple to  fields living on higher branes that encompass
our four-dimensional worldvolume.

Local tadpole cancellation constrains severely model building, and in
particular the mechanisms of supersymmetry breaking.
The case of uncancelled local tadpoles in
$d_\perp =2$, on the other hand,  appears particularly interesting  because of the
logarithmic sensitivity of the Standard Model parameters on the size of the
transverse space. This leaves open the possibility of dynamically determining
the hierarchy by minimizing an effective potential on our
world-brane. It is to this end crucial to 
know how to resum all the
large logs -- as  in renormalizable field theory where they
are absorbed in  a finite number of renormalizable
couplings. This may sound hopeless in our context, since there is no
field-theory description beyond the TeV.
 Fortunately, we will argue that an  equally-powerfull  procedure is  available:
all large logs can be  effectively
absorbed into the values of finitely-many  massless  fields living in
the bulk and evaluated at the (transverse) position of our world-brane. 
 The role of the renormalization
group equations is thus  played by the
 classical supergravity equations in the effectively two-dimensional
transverse space -- with 
higher derivative terms being  ignored because the variations of
fields are  logarithmic.

It is important to  appreciate  that in a non-supersymmetric string
theory the  bulk one-loop vacuum energy, 
$\Lambda_{\rm bulk}\sim M_s^{10}$,  gives rise  to a four-dimensional
cosmological constant with quadratic sensitivity on the Planck scale,
$\Lambda\sim M_P^2 M_s^2$ {\adpq}. This is analogous to the
 quadratically-divergent one-loop
contribution, proportional to Str${\cal M}^2$, in a softly-broken 
supersymmetric field theory.
 Since the effective potential on the
world-brane has at best  logarithmic sensitivity on $M_P$, 
the hierarchy might
be determined dynamically only if such quadratic corrections  are absent
\str, \adpq. 
In the framework of softly broken supersymmetry  this
condition appears to be  adhoc, unless the breaking is induced by
compactification in a way analogous to turning on  
finite temperature \ia \tfour. In the novel framework, on the other
hand, the spontaneous
breaking can  be confined to the world-brane at the string scale, 
and then communicate very weakly to  the bulk \aadd. This can for
instance be achieved by turning on internal magnetic fields, or in
T-dual language by rotating the branes in the compact space \mag\doug.

\bigskip
\break

{\centerline{\bf Ultraviolet versus Infrared}}
 \bigskip

If all the Standard Model fields live on a  set of
D-branes, then the observed non-gravitational processes are given by
amplitudes whose external legs are open strings with endpoints on these
D-branes. The branes must of course extend along the four dimensions
of our physical space-time. We assume for simplicity that the only
other dimensions whose  size is significantly-larger than the string length
are transverse to the common worldvolume of our branes.
The possible soft (substringy) momenta in the theory  have thus a
four-dimensional longitudinal component ($p_\parallel$) and a transverse
component
($p_\perp$). The quarks, leptons and gauge bosons of the Standard
Model, i.e. all  external legs in the amplitudes of interest,
 are not allowed to  carry any transverse momentum.


\vskip 0.8cm

\centerline{\psfig {figure=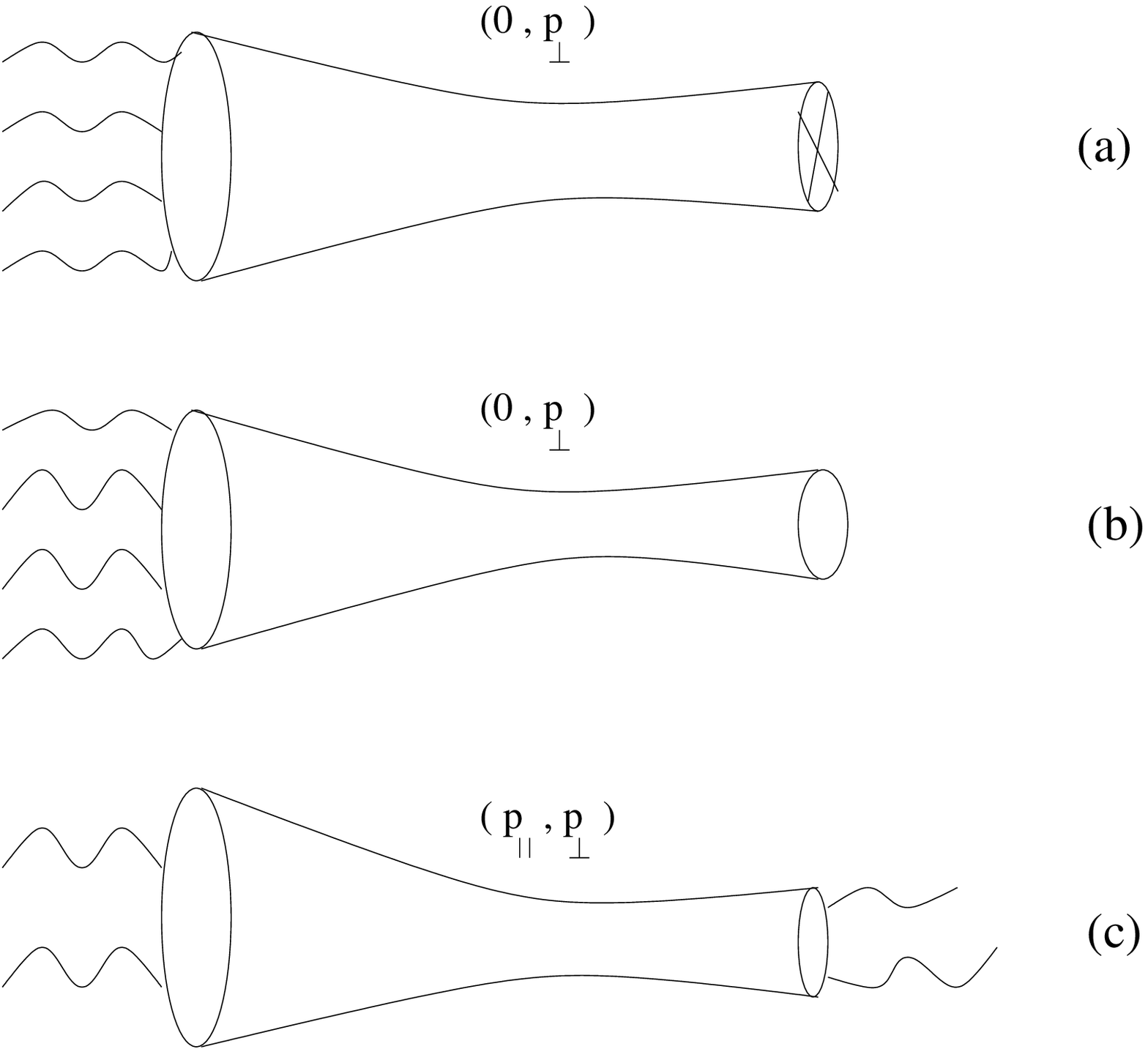,height=4truein}}
\vskip 0.6cm
\noindent{\ninepoint\sl \baselineskip=8pt {\bf Fig.1}: {\rm
The (a) M{\"o}bius, (b) planar annulus, and (c) non-planar annulus
diagrams contributing to a 4-point  amplitude of world-brane fields.
On top of every diagram we indicate the momentum  flowing along  the cylinder.
The infrared divergences of diagrams (a) and (b) appear as
ultraviolet quantum-gravity effects  to an observer localized on the
world-brane.
}}

\bigskip

Consider now the  one-loop  contributions  to such amplitudes, 
 given by the  three  diagrams
of  figure  1~: (a) M{\"o}bius,  (b) planar
annulus with all external legs on the same boundary, and (c) non-planar
annulus with some external legs on each of the  two  boundaries.
All  potential divergences of these amplitudes are interpreted
by  the ten-dimensional string theorist
as large infrared effects.  They occur because
some massless (or very light) intermediate
open- or closed-string  state  propagates  very nearly on-shell. The
four-dimensional Standard-Model physicist,  on the other hand, may
interpret these effects very differently.  Some of them will look as genuine
infrared effects,  already incorporated in his  effective low-energy action \sm,
while others modify his  parameters at the weak scale  and appear
intimately related  to  the ultraviolet structure  of the fundamental theory!

Let us take a closer look at these  potential divergences. Those
 associated with light open strings --  the photon and gluons in
 particular -- occur at {\it exceptional} values of the external
 momenta  and are the usual infrared effects of
 Yang-Mills theory. Likewise, the potential divergence  in  the non-planar
 amplitude (c) is to the eyes of our brane physicist  a soft
 effect coming from  the exchange of a   gravitational particle. Indeed,
for generic momenta of the external legs,  
the  intermediate closed string in the transverse channel
carries  non-vanishing four-momentum $p_\parallel$, which  cuts off
 the putative infrared divergence.\foot{Note that $p_\parallel$ is
 conserved at each vertex, while $p_\perp$ is not conserved since
 the branes break the translation invariance in transverse
space.} Strictly-speaking the effective theory {\sm}  does not,
of course,  
account correctly for the fact that  gravity is 
higher-dimensional above the Kaluza-Klein  scale $1/R$, with $R$ the
typical radius of the large transverse dimensions. For instance, for a
 single large dimension the tree-level exchange amplitude   is of the order of
\eqn\exch{
{\cal A} \sim \left( {p_\parallel\over M_s}\right)^{2n} {\cal
G}^{(2)}\ ,
\ \ {\rm where}\ \ \  {\cal
G}^{(2)} \sim 
{1\over p_{\parallel}}{\rm coth} (\pi p_{\parallel} R)\
}
is the two-point function on the brane   obtained
 by summing over the Kaluza-Klein modes,
 $p_{\parallel}$ denotes here the modulus  of the longitudinal
momentum, and the vertex-suppression factor in the amplitude  has
$n=2$ or $1$
for gravitational or gauge-like couplings.
 For ultrasoft momenta $p_\parallel\ll 1/R$,  the two-point function   ${\cal
G}^{(2)} \sim {1/ \pi R
p_\parallel^2} + \pi R/3$ exhibits 
 the standard inverse-square  behaviour plus a large constant
correction due to the extra dimension {\karim}. For 
$p_\parallel\gg  1/R$ on the other hand,   
 ${\cal
G}^{(2)}\sim 1/ p_\parallel$ 
corresponding to  Newton's law in five dimensions. The basic point for
our purposes here, in any case,  is
that at current accelerator energies  and for 
 non-exceptional values of momenta   such 
gravitational  exchanges  remain  suppressed,  and all  potential  infrared
divergences are  regularized.

The M{\"o}bius and the planar-annulus diagrams, on the other hand,
are very different. Since there is no
4d momentum flowing down  the cylinder,  their potential divergences
 persist  for arbitrary external momenta at  the weak scale. Such
 large effects will thus  correct the tree-level parameters of the
 Standard Model, and must be kept under control to avoid
 destabilizing  the gauge hierarchy. 
  In order for such effects to be present, a massless (or light)
  closed string  with soft transverse momentum $p_\perp \ll M_s$
 should be allowed to disappear into the vacuum. This can only happen
  if there are local tadpoles  uncancelled at  distances
  large compared to the string scale,
  $l\sim 1/p_\perp\gg l_s$.  Global ($p_\perp = 0$) tadpole
  cancellation in a
  compact space is of course  required for  topological consistency -- our
  perturbation theory would otherwise be truly  divergent and hence
  sick.  Local tadpoles,
 on the other hand, need not a priori destroy perturbativity,
and  their effects can as we will argue  be effectively resummed.
 The simplest context in which to discuss these ideas is type-I or type-I'
theory   with  $2^d$ orientifold
 $(9-d)$-planes transverse to a $d$-dimensional torus $T^d$.
 Global consistency requires also 32
 D$(9-d)$-branes, which are free classically to  move at arbitrary
  positions on the torus. Tadpoles will
 cancel locally only at one special point of  moduli space --
when the 32 D-branes  are
  split in  groups of $2^{5-d}$, with each group  sitting
 precisely  at an  orientifold plane \polwit \bgs   \ads. Another simple example is the
   $T^4/Z_2\times T^d$ orbifold  \bs, in which only  $T^d$  is
considerably   larger than string size. Global tadpole cancellation requires 32
 D$(9-d)$-branes and $32$
 D$(5-d)$-branes, all free to move on the transverse torus. Local
  tadpole cancellation on $T^d$ is again achieved at a special point
  in moduli space, i.e. when the D-branes  split in equal numbers,
  with each group sitting at one of the  $2^d$ orientifolds \berkooz\ads.

The contribution of these local  tadpoles to the world-brane
  amplitudes can be estimated easily as follows.
 Assuming  $d$ large transverse dimensions
$x^i\in [0, 2\pi R_i]$   we find the following
 estimate for what looks to a brane physicist as a large
 `ultraviolet' contribution:
\eqn\div{
{\rm uv}({\cal A}) \ \
\sim\ \  {1\over V_\perp}\  \sum_{\vert {p}_\perp\vert < M_{s} }\
{1\over p_\perp^2}\  F({\vec p_\perp})\, ,
}
where $V_\perp= \prod_i R_i$ is the volume of the transverse space,
 ${\vec p}_\perp = ( {n_1/ R_1}, \cdots , {n_{d}/ R_{d}})$
is the  transverse momentum carried away by the massless closed string, and
$F({\vec p_\perp})$ are the local tadpoles,
 Fourier-transformed to  lattice-momentum space. The
tadpoles arise from the distribution of  the D-branes and the orientifolds
which act as classical point-like  sources in the transverse space.
In the specific examples  discussed above, the  tadpoles have the
generic form
\eqn\tad{
F({\vec p_\perp}) \sim   \left( 2^{5-d} \prod_{i=1}^{d} \left(1+ (-)^{n_i}\right)
 - 2  \sum_{a=1}^{16} {\rm cos}( {\vec p_\perp}{\vec x_a})  \right)\, ,
}
where the orientifolds are located at the corners of the cell $\prod_i
 [0, \pi R_i]$ , and 
$\pm {\vec x_a}$
are  the transverse positions  of the 32  D-branes. These positions
 correspond to the Wilson lines of  the T-dual picture.
The momentum sum in equation {\div}  is cutoff effectively at $M_s$ because of
the  form-factor
suppressing exponentially the amplitude at higher scales. Heuristically,
this is because the external  open strings are localized on the
world-brane with  an uncertainty
 $o(l_s)$ --  this  is  indeed the size of a D-brane as seen by a
fundamental-string probe \kleb.

 For generic positions of the D-branes, and for roughly equal radii
 $R_i \sim R \gg l_s$,
the above expression has the following behaviour in the
decompactification limit:
\eqn\beh{
{\rm uv}({\cal A}) \sim  \cases{ o(R)\ &for \ $d=1$\cr
o($log$R)\ &for \ $d=2$\cr
${\rm finite}$ \ &for\ \ $d>2$\cr}
}
This result is clearly  dictated by the large-distance behaviour
of the two-point  function in the $d$-dimensional transverse space.
The conclusion is that the radius $R$, and hence
also the four-dimensional Planck scale
 $M_P\sim M_s (M_sR)^{d/2} $, decouples
from the loop corrections to  world-brane parameters  for $d >2$
large transverse dimensions. It also of course decouples for $d=1,2$
if the tadpoles cancel exactly locally.
 In either case the technical
problem of the gauge hierarchy is solved, but there is little room for
understanding its dynamical origin from the world-brane viewpoint.
\foot{Proposals for stabilizing the hierarchy 
were discussed in refs. \stab.}
 When  $d=1$ there
are large linear corrections and,  as we will explain  in a minute,  one hits
quickly a strong-coupling singularity forbidding the expansion of the
transverse space. Thus, the marginal case $d=2$ is singled out
 as the most promising candidate
for generating and solving the hierarchy.

   Our discussion can be generalized easily to anisotropic
compactifications, and/or in the presence of higher branes which extend
partially into the ten-dimensional bulk. 
The precise  criterion for
solving the technical gauge-hierarchy problem is that all massless
bulk fields coupling to our Standard Model have long-distance Green  functions
 in  the transverse space
which are  at most logarithmically large. 
We may define  an `effective dimensionality' $d_\perp$ in
which the bulk fields with uncancelled local tadpoles propagate. 
 For fields
confined for example   to a $3+p$-brane,  $d_\perp$ is at most equal to
$p$. If the transverse space is anisotropic, with $n$ large
dimensions of size $r$   and one even  larger of size $R$, then the Green
function would exhibit linear growth when $R\gg r^{n}$ in string units, in which
case the effective dimensionality  is $d_\perp =1$. Such possibilities
have been  discussed in ref. {\cb}. 
The condition for solving the technical problem associated with the 
gauge hierarchy is  that for all massless bulk fields with uncancelled
tadpoles $d_\perp \ge 2$. Logarithmic sensitivity is possible even if
all six dimensions become large, provided the effective dimensionality
of some bulk fields is two.

\vskip 0.5cm

{\centerline{\bf The New Guise of the Renormalization Group}}
\bigskip

According to the
traditional viewpoint the physics between the electroweak and Planck
scales is described by renormalizable four-dimensional field
theory. This means that all large quantum corrections involving
ultraviolet degrees of freedom can be absorbed into a finite number of
parameters -- the masses and renormalizable couplings of the Standard Model
 measured at the electroweak scale $M_Z$.
If the theory beyond $M_Z$ is a string theory, this traditional logic
is not valid. Nevertheless, as we will now explain, in the particular
case
$d_\perp =2$ all large quantum
corrections involving the transverse volume, and through it the
four-dimensional Planck scale, can be similarly  absorbed into  a
 finite number of parameters of the brane theory. These parameters
correspond to the values of the massless bulk background fields on the
four-dimensional world-brane. The classical bulk equations of
supergravity play the role of  the renormalization group equations, in
that they allow a resummation of the leading logarithmic corrections.

 Let us  explain this important point in some more detail, in the
 context of toroidal  compactifications of type-I theory. The classical
 background has constant dilaton, metric and Ramond-Ramond (RR) gauge fields,
as well as a number of $\delta$-function sources representing the D-branes
 and the orientifolds. The number of these sources is being fixed by
 global  tadpole cancellation --  or else  compactification
 of the transverse space  would be  obstructed  topologically.
 Tadpoles need not however cancel locally, since the D-branes can be
 placed at arbitrary positions. These local sources modify the
 background fields, an  effect that is  classically invisible because
 the strength of the sources  is controlled by the string coupling
 constant $g_s$. In the full quantum theory, on the other hand, we
 have to take into account the modification of the backgrounds.

Consider now a very large (in string units) transverse space, and
    focus on some particular set of D-branes on which live our
    Standard Model gauge and matter fields. Their effective Lagrangian
    depends on the dynamics of the bulk only through the values of the
    supergravity backgrounds, evaluated  at the position of these
    D-branes. If the number of large transverse dimensions
 $d_\perp > 2$,  the effects  of the distant sources die  out  and the bulk
    fields can be approximated by their constant classical
    background values. The  size and other macroscopic moduli of the transverse
    space decouple from  the quantum-corrected   brane theory,
    in the same way that the ultraviolet cutoff decouples from a
    {\it super-renormalizable}  field theory.
 Examples of such models with $d_\perp > 2$
can be provided by T-duals of type-I compactifications
 with only D9-branes, in which
$d_\perp$ can reach its maximum value of six, or
 with at most one set of D5-branes in which case
 the maximum $d_\perp$ equals four.

In the case $d_\perp =2$ the effect of the distant sources grows
at most  logarithmically with size, so that the gradients of bulk fields
away from these sources stay  small.
It can be argued that the classical supergravity equations in
the transverse space can be used to resum all large corrections by
solving for the spatial variation of the background fields.
Indeed, $\alpha^\prime$ corrections are negligible because the
gradients of the background fields stay small.
In what concerns supergravity (closed string) loops, these are higher dimensional
and hence, by power counting, lead to higher derivative corrections
which are suppressed. An independent argument follows from
supersymmetry in the bulk which is at least six dimensional; 
this ensures that the two-derivative supergravity action receives no
corrections. As a result,
the quantum-corrected Lagrangian of the brane theory is obtained by 
evaluating these bulk  backgrounds  at the position
of our world-brane. Note that this position is only defined within
$\sim o(l_s)$,  since at shorter distances the classical supergravity
equations don't make sense \size. The main message is that when $d_\perp =2$
we have  a  method for resumming the large logs which is
equally-powerful as the renormalization-group equations in the case of
{\it renormalizable} quantum field theory.

In the final case $d_\perp =1$ the effect of the distant sources grows
linearly,  and  one  expects in general  that
some  brane theory will become  strongly coupled very rapidly
as  the value of 
the radius is being increased beyond string length. 
For instance in semi-realistic  compactifications with non-maximal supersymmetry
($N=2$ or $N=1$), gauge couplings will in general acquire large threshold
corrections, that grow  linearly with the radius. If for some gauge group factor
these are  negative, the corresponding gauge coupling will hit a
strong coupling singularity  soon after  the radius
moves away from the string length. This leads to an upper
 bound for the size  of the
transverse dimension, similar to the one obtained in the compactification of
M-theory on Calabi-Yau$\times S^1/Z_2$ with standard embedding \w.
The phenomenon
can  be also understood  as a consequence of the non-factorization of the internal
manifold, due to a position dependence of the Calabi-Yau
volume along the line-segment. One can of course avoid such
singularities if tadpoles cancel exactly locally, or if there existed 
some special models in which for instance 
all linear thresholds happen to be positive, so that
gauge couplings  become infinitesimal in the large-radius limit. Such
cases are however exceptional and too restrictive for model building.
In the generic case bulk-field  variations cannot be controlled by the
supergravity equations, since in particular quantum gravity effects
will become at least locally strong. The situation is therefore here
analogous to that of {\it non-renormalizable} field theories.

To summarize our conclusions in this section, we draw below  a schematic
correspondence between  the dependence on the ultraviolet cutoff
in quantum field theory, and the dependence on $M_P$ in the brane picture.

\vskip 0.5cm

\hskip 70pt
\vbox {\tabskip=0pt \offinterlineskip\def\tablerule{\noalign{\hrule}} 
\halign to200pt {\strut#& \vrule#\tabskip=1em plus2em& \hfil#& \vrule#& 
 \hfil#& \vrule# \tabskip=0pt\cr\tablerule
  &&\omit\hidewidth $d_{\perp}$ \hidewidth&&
 \omit\hidewidth QFT analog \hidewidth&\cr\tablerule
&&1\ \ &&non-renormalizable &\cr\tablerule
&&2\ \ &&renormalizable\ \ \  &\cr\tablerule
&& $>$2  &&super-renormalizable &\cr\tablerule
\hfil\cr}}

\vskip 0.5cm

An implication of our discussion is that the effective field-theory
couplings of bulk fields to the brane {\coupl}  are  generally
modified. Firstly  there can be bulk fields other than the graviton which
can be emitted in the transverse space. Secondly, the local strength
of the coupling on the brane can be enhanced or reduced relative to
the four-dimensional Planck
scale. As a result the bounds on the transverse size and the string scale
obtained from exotic events with missing
energy {\giu} may  need to be further refined.

\vskip 0.5cm
\break

{\centerline{\bf Two Examples}}
\bigskip

We will conclude  our discussion  with two simple
examples that will help illustrate the main points of this letter.
Consider first the
$T^4/Z_2\times T^2$ orbifold of type I theory, whose threshold
corrections to the 4d gauge couplings
were analyzed in refs. \cyril  \cb. We take $T^2$  large and $T^4/Z_2$
of order the string size, and will use the type-I'
description in which one has 32 D3-branes and 32 D7-branes, all
transverse to the large $T^2$. Local tadpole cancellation in the large
transverse space requires that exactly eight D3 branes and eight
D7-branes are located in the vicinity (within $\sim l_s$) of each of the
four orientifold planes. Our observable world would correspond to one
such set.
In this case the radii of $T^2$ decouple from quantum corrections
to our  effective world-brane action. This implies in particular that
the threshold corrections to the
4d gauge couplings, at this point in moduli space,
should not depend on the geometry of $T^2$.

To check this note first that because of the N=2 supersymmetry
only charged BPS states contribute to the renormalization of the gauge
couplings \cyril. The BPS open-string states can be most easily read off at
the point of maximal $U(16)\times U(16)$ gauge symmetry, corresponding
to all D-branes sitting at the same  orientifold. Besides the adjoint
vector multiplets, the massless  states at this point are  hypermultiplets
in the  antisymmetric ${\bf 120}$ and $\overline {\bf 120}$ representation of
each gauge
group, as well as a hypermultiplet in the representation
$({\bf 16},{\bf 16})$.  The BPS states include all winding excitations
on the two-torus. In the configuration in which the tadpoles cancel
locally, the gauge group at each orientifold is $U(4)\times U(4)$,
with massless hypermultiplets in the ${\bf 6}$ and $\overline {\bf 6}$
for each group factor, and one hypermultiplet in the $({\bf 4},{\bf
4})$. Since all other states are hypermassive ($\sim R_i M_s^2$),
our world-brane theory can decouple from the size of the torus only if
its $\beta$-functions vanish identically. The $\beta$-function
contributions to a $N=2$ $SU(n)$  theory are one-loop and proportional to
\eqn\bet{
b_{SU(n)}= \cases{\ \ 2n \ \ \ \ \ {\rm adjoint} \cr
n-2\ \ \ {\rm antisymmetric}\cr
\ \ \ 1\ \ \ \ \ \ {\rm fundamental}\cr}  
} 
It is thus straightforward to check that they vanish for the $SU(4)^2$
parts of the gauge group, thus confirming that the large logarithmic
corrections are absent when the tadpoles cancel locally.
Notice that the $U(1)^2$ factors become massive by
mixing with bulk tensor hypermultiplets
\berkooz, so their apparent non-vanishing $\beta$-functions  can be
understood through the exchange of this hypermultiplet.

   With the  second example, we want to illustrate the general claim
   that all large effects in the decompactification limit of a 2$d$
   transverse space can be absorbed in a finite number of parameters,
   which correspond to the values of the bulk fields at the position
   of our world-brane. These fields have no longitudinal
   space dependence, and are therefore  parameters of the brane theory.
The simplest case is that of a maximally supersymmetric
compactification of type-I theory on a two-torus. As before, we
take $T^2$ very large and use type I' language, so that there are
32 D7-branes and four orientifolds, all
transverse to the two-torus. The world-brane action of the gauge fields
  in units of the fundamental string tension ($T_F=1$) reads {\revs}
\eqn\tree{
{\cal L}_{brane} = {1\over 4} T_{(7)}^I\;
\sqrt{g}\;  e^{-\phi} {\rm tr} \left(  F^2 -{1\over
48}  t_8 F^4 + \cdots
+ 2  C^{(4)} F\wedge F + {1\over 6} C^{(0)} \epsilon_8 F^4 \right)
}
where $ T_{(7)}^I$ is the D7-brane tension,
 $g$ is the induced metric on the worldvolume,
$t_8$ is the usual eight-index tensor that appears in the
expansion of the Born-Infeld Lagrangian, $\epsilon_8$ the
totally-antisymmetric tensor,
$\phi$ is the ten-dimensional
dilaton, and $C^{(n)}$ the RR antisymmetric $n$-form fields.
Assuming eight-dimensional Poincar\'e invariance sets $C^{(4)}=0$. The
remaining
 bulk fields $\phi$ and $C^{(0)}$, together with the scale factor
$\omega$ of the 8d worldvolume flat space
${\bf R}^8$ and the three-component metric $G_{ij}$ of the transverse space
can be functions of the transverse coordinates $\xi^i$, but constant along
${\bf R}^8$. Note that the ten-dimensional metric in the string frame used
above reads
\eqn\met{
ds^2 = \omega^2 dx^\mu dx^\nu \eta_{\mu\nu} + G_{ij} d\xi^i d\xi^j\ .
}
These functions are non-trivial except at the special point in
moduli space where there are eight D7-branes at each orientifold, so
that all tadpoles cancel locally. The precise functions can be read
off by considering this background as a compactification of F-theory
on an elliptically-fibered K3  \F. The transverse space is the
base-space,  and $\tau = C^{(0)} + i e^{-\phi}$ is the complex
structure of the torus fiber.

   Without describing in detail these variations, let us just note
   that by evaluating these bulk fields at the position of our
   world-brane we can find the quantum-corrected parameters of the
   brane theory, for any given point in moduli space. Consider, in
   particular, the CP-even part of the gauge-field  action, which
   reads
\eqn\BI{
{\cal L}_{brane} = {1\over 4}T_{(7)}^I\;
  e^{-\phi} {\rm tr} \left( \omega^4  F^2
 -{1\over 48} t_8
 F^4 + \cdots \right)\, .
}
To find the variation of the fields, one must consider the bulk
 supergravity equations reduced from ten down to the two transverse
 dimensions. The relevant action is
\eqn\two{
S \propto \int d^2\xi  \sqrt{G} e^{-2\phi_{(2)}}\left( {1\over 2} {\cal
R}^{(2)} - 2
(\partial\phi_{(2)})^2
+4  {(\partial \omega)^2 \over \omega^2}  + {e^{\phi_{(2)}}\omega^4
\over 16\pi^3 \sqrt{G}}
\sum_a \delta^{(2)}(\xi-\xi_a) \right)
}
where $\phi_{(2)}$ is the two-dimensional dilaton, $e^{-\phi_{(2)}}=
e^{-\phi}\omega^4$,  while ${\cal R}^{(2)}$ is the 2d scalar curvature, and the
$\delta$-function sources correspond to the D7-branes and the
orientifolds.

 Let us for instance explain in this context why the
$F^2$ term in {\BI} receives no quantum corrections -- either perturbative or
non-perturbative. The coefficient of this term
 is precisely the exponential of the
two-dimensional dilaton, which satisfies a free-field equation without source
obtained by varying the conformal factor of the two-dimensional metric
in {\two}. As a result $\phi_{(2)}$ is
 a constant in transverse space, equal to its
classical  value, and 
 independent  of the positions
of the D-branes as well as  of the geometric torus moduli. This can be
checked alternatively from the explicit form of the seven-brane
solution {\seven}  in which $e^{-\phi}\omega^4$ can be seen to be constant.
The coefficient $e^{-\phi}$ of the $F^4$ term,  on the other hand, has
a variation which can be read off the F-theory solution. 
When  one perturbs around the special configuration
in which  tadpoles cancel locally, one finds \F
\eqn\sen{
e^{-\phi} = 1  -{ g_s\over 2\pi} \sum_{a=1}^4
 {\rm log}\vert 1- z_a/z\vert
}
where $z= \xi^1+i\xi^2$ is a complex coordinate, $z=0$ is the position
of the orientifold and $z_a$ are the positions of the (pairs of)
D7-branes. To evaluate this expression on the $b$th brane one
must set $z = z_b + o(l_s)$, where $l_s$ plays the role of the
ultraviolet cutoff analogous to the one imposed in equation {\div}. It is
easy to see that these logarithms, which arise from the two-dimensional
Green's function, reproduce the one-loop logarithmic dependence  of the $F^4$
term in this context {\ff}. Note that the splitting of the orientifold
 planes demonstrated by Sen {\F}  affects our discussion by
 corrections that are exponentially-supressed at weak coupling.

\vskip 0.5cm

{\bf Acknowledgments}:
I.A. would like to thank the Physics Department of Stanford University
and C.B. the Aspen Center for Physics for hospitality during initial
stages of this work. We have benefited from discussions with K. Benakli,
S. Dimopoulos, M. Green and A. Kumar.
This work was partially supported by the EEC grant  TMR-ERBFMRXCT96-0090.

\listrefs

 \bye